\documentclass[showpacs,showkeys,amssymb,preprintnumbers,nofootinbib,
superscriptaddress]{revtex4}

\usepackage{amsmath}
\usepackage{graphicx}
\usepackage{latexsym}
\usepackage{color}
\usepackage{amsfonts}
\usepackage{url,hyperref}
\usepackage{bm}

\newcommand{\nn}{\nonumber\\}
\newcommand{\beq}{\begin{equation}}
\newcommand{\eeq}{\end{equation}}
\def\bea{\begin{eqnarray}}
\def\eea{\end{eqnarray}}

\begin{document}

\title{Black hole quasinormal modes using the asymptotic iteration method}
\author{H.~T.~Cho}
\email[Email: ]{htcho@mail.tku.edu.tw}
\affiliation{Department of Physics, Tamkang University, Tamsui, Taipei, Taiwan, Republic of 
China}
\author{A.~S.~Cornell}
\email[Email: ]{alan.cornell@wits.ac.za}
\affiliation{National Institute for Theoretical Physics; School of Physics, University of the 
Witwatersrand, Wits 2050, South Africa}
\author{Jason~Doukas}
\email[Email: ]{jasonad@yukawa.kyoto-u.ac.jp}
\affiliation{Yukawa Institute for Theoretical Physics, Kyoto University, Kyoto, 606-8502, Japan}
\author{Wade~Naylor}
\email[Email: ]{naylor@eml.doshisha.ac.jp}
\affiliation{Department of Electrical Engineering, Doshisha University, Kyotanabe, Kyoto, 510-3210, Japan}
\affiliation{Department of Physics, Osaka University, Toyonaka, Osaka 560-0043, Japan}
\thanks{Main address from 1st of June.}

\begin{abstract}
  In this article we show that the asymptotic iteration method (AIM) allows one to \textcolor{black}{numerically} find the quasinormal modes of Schwarzschild and Schwarzschild de Sitter (SdS) black holes. An added benefit of the method is that \textcolor{black}{it can also be used to calculate} the Schwarzschild anti-de Sitter (SAdS) \textcolor{black}{quasinormal modes} for the case of spin zero perturbations. We also discuss an improved version of the AIM, more suitable for numerical implementation.
\end{abstract}

\date{May, 2010}
\preprint{YITP-09-110, WITS-CTP-049}
\maketitle

\section{Introduction}
\label{intro}

\par The study of quasinormal modes (QNMs) of the Schwarzschild black hole is an old and 
well established subject (for a recent review see reference \cite{Berti:2009kk} and references 
therein), where the various frequencies have been well determined, typically by applying a 
Frobenius series solution approach, leading to continued fractions for QNM boundary conditions, \'a la Leaver \cite{Leaver:1985ax}. Recently a new method for obtaining analytic/numerical solutions of second order ordinary differential equations with bound potentials has been developed called the 
asymptotic iteration method (AIM) \cite{Ciftci:2003}, which was found to be closely connected 
to continued fractions developed from exact WKB solutions, see reference \cite{Matamala:2007} and those listed in therein.\footnote{The  phrase ``continued fraction" used for WKB solutions should not be confused with the type of continued fractions developed from Frobenius series for QNMs by Leaver \cite{Leaver:1985ax}. In the rest of the paper ``continued fraction'' will mean those generated from Frobenius series.} 

\par In this article we will demonstrate that the AIM can also be applied to the case of 
black hole QNMs, which have unbounded (scattering) like potentials. On a related note, the 
AIM was used to find QNMs for Scarf II (upside-down Poschl-Teller-like) potentials \cite
{OzerRoy}, based on observations made by one of the current authors \cite{ChoLin} relating 
QNMs from quasi-exactly solvable models. Indeed bound state Poschl-Teller potentials have 
been used for QNM approximations previously by inverting black hole potentials
 \cite{Ferrari:1984zz}. However, the AIM does not require any inversion of the black hole potential as we shall show.
 
\par In this paper we shall focus on spherically symmetric backgrounds in four dimensions with field equations of the form:
\beq
\frac{d^2 \psi(x)}{dx^2} + \left[ \omega^2 - V(x) \right] \psi(x) = 0 ~,
\label{radtort} 
\eeq
where $V(x)$ is a master potential of the form \cite{Berti:2009kk}
\beq
V(\textcolor{black}{r}) = f(r) \left[ \frac{\ell (\ell +1)}{r^2} +(1-s^2)\left( \frac{2M}{r^3} - {(4-s^2) \Lambda\over 6} \right)\right]~,
\label{mastpot}
\eeq
\textcolor{black}{and $dx=dr/f(r)$ where}
\beq
f(r)= 1- {2M\over r} -{\Lambda\over 3}r^2
\eeq
\textcolor{black}{with cosmological constant $\Lambda$}. Here $s=0,1,2$ denotes the spin of the perturbation: scalar, electromagnetic and gravitational, (for half-integer spin see references 
\cite{Zerilli:1971wd,Medved:2003rga,Cho:2003qe}). In the following we shall use the definition that QNMs are defined as solutions of the above 
equations with boundary conditions: 
\beq
\psi(x) \to \left\{
\begin{array}{cl}
e^{i \omega x}   &\qquad x\to \infty \\
e^{-i \omega x}  &\qquad x\to -\infty ~,
\end{array}
\right.
\label{QNMdef}
\eeq
 for $e^{-i \omega t}$ time dependence (which corresponds to ingoing waves at the horizon and outgoing waves at infinity). We shall see how the AIM can not only be applied to Schwarzschild and Schwarzschild de Sitter (SdS) QNMs for general spin, $s$, but also for the Schwarzschild anti-de Sitter case (SAdS), at least for spin zero ($s=0$) fields. 
 
\par The structure of the paper is as follows. In the next section we describe how the AIM can be applied to exterior eigenvalue problems for the Schwarzschild background. Then in Section \ref{extra} we show how the AIM can be applied to backgrounds with more than one horizon,  such as SdS. In Section \ref{extraA} we show how the AIM can be applied to scalar fields on SAdS, while in Section \ref{results} we discuss the results and make conclusions. In an appendix we present a brief derivation of the continued fraction method for SdS for spin $s=1,2$ fields.

\section{Schwarzschild Black Holes}
\label{schwarz}

\par To explain the asymptotic iteration method (AIM) we shall start with the simplest case of the radial component of a perturbation of the Schwarzschild metric outside the event horizon \cite{Zerilli:1971wd}. For an asymptotically flat Schwarzschild solution ($\Lambda=0$):
\beq
f(r)= 1- {2M\over r} \,\,\, , 
\eeq
where from $dx = dr/f(r)$ we have 
\beq
x(r)=r + 2M \ln\left({r\over 2M}-1\right)\,\,\, , 
\eeq
for the tortoise coordinate, $x$. Note, for  the Schwarzschild background the maximum for this potential, in terms of $r$, is given by 
\beq
r_0= {3M\over 2} {1\over \ell (\ell+1)} \Big( \ell(\ell
+1) - \sigma + (\ell^2(\ell+1)^2 + {14\over 9} \ell(\ell+1)\sigma +\sigma^2)^{1/2}\Big)~,
\eeq 
where $\sigma=1-s^2$ \cite{Iyer:1986nq}. 

\par The choice of coordinates is somewhat arbitrary and in the next section (for SdS) we will see how an alternative choice leads to a simpler solution. First, consider the change of variable: 
\beq
\xi =1 - \frac{2 M}{r} ~,
\eeq
\textcolor{black}{ with $0 \leq \xi < 1$}.  \textcolor{black}{In terms of $\xi$, equation (\ref{radtort}) 
then becomes}, 
\beq
\frac{d^2 \psi}{d\xi^2} + \frac{1 - 3 \xi}{\xi (1 - \xi)} \frac{d \psi}{d\xi} + \left[ \frac{4 
M^2\omega^2}{\xi^2 (1 - \xi)^4} - \frac{\ell (\ell + 1)}{\xi (1 - \xi)^2}  -\frac{1-s^2}{\xi (1-
\xi)} \right] \psi = 0~. 
\eeq

\par To accommodate the out-going wave boundary condition $\psi \to e^{i\omega x}=e^{i
\omega(r+2M \ln(r/2M-1))}$ as $(x,r)\to \infty$ in terms of $\xi$ (which is the limit $\xi \to 
1$) and the regular singularity at the event horizon ($\xi\to 0$), we define 
\beq
\psi (\textcolor{black}{\xi})= 
 \xi^{- 2iM\omega}  (1-\xi)^{-2iM\omega} e^{\frac{2 i M \omega}{1 - \xi} }\chi (\textcolor{black}{\xi}) ~,
\eeq
where the Coulomb power law is included in the asymptotic behaviour  (cf. reference \cite
{Leaver:1985ax} equation (5)). The radial equation then takes the form:
\bea
\chi''& = & \lambda_0(\xi)  \chi' + s_0(\xi) \chi~ ,
\label{AIMform}
\eea
where
\bea 
\lambda_0(\xi) & = & \frac{4M i \omega (2\xi^2 - 4 \xi + 1) - ( 1 - 3 \xi)(1 - \xi)}{\xi (1 - \xi)
^2} \,\,\, , \\
s_0 (\xi)& = &  \frac{16M^2  \omega^2(\xi - 2) - 8M i \omega ( 1 - \xi)+\ell (\ell + 1) + 
(1-s^2)(1 - \xi)}{\xi (1 - \xi)^2} ~.
\eea
Note that primes of $\chi$ denote derivatives with respect to $\xi$. 

\par The crucial observation in the AIM is that differentiating the above equation $n$ times with 
respect to $\xi$, leaves a symmetric form for the right hand side:
\beq\chi^{(n+2)} = \lambda_n(\xi) \chi' + s_n(\xi) \chi \,\,\, , 
\eeq
where
\beq \label{eqn:nthiter}
\lambda_n(\xi) = \lambda'_{n-1} (\xi)+ s_{n-1}(\xi) + \lambda_0(\xi) \lambda_{n-1}(\xi) 
\hspace{1cm} \mathrm{and} \hspace{1cm} s_n(\xi) = s'_{n-1}(\xi) + s_0(\xi) \lambda_{n-1}
(\xi) \,\,\, . 
\eeq
For sufficiently large $n$ the asymptotic aspect of the ``method" is introduced, that is:
\beq
\frac{s_n (\xi)}{\lambda_n (\xi)} = \frac{s_{n-1}(\xi)}{\lambda_{n-1}(\xi)} \equiv \beta(\xi) 
\,\,\, , 
\eeq
where the quasinormal modes are obtained from the ``quantization condition" 
\beq\label{eqn:quantcond}
\delta_n = s_n \lambda_{n-1} - s_{n-1} \lambda_n =0~,
\eeq  
which is equivalent to imposing a termination to the number of iterations \cite{Barakat:2006}.
This leads to the general solution
\beq
\chi(\xi) = \exp \left[ - \int^\xi \beta (\xi') d\,\xi' \right] \left( C_2 + C_1 \int^\xi \exp \left
\{ \int^{\xi'} \left[ \lambda_0 (\xi'') + 2 \beta ( \xi'') \right] d\,\xi'' \right\} d\,\xi' \right) \,\, 
\eeq
for integration constants $C_1$ and $C_2$, which can be determined by an appropriate choice of normalisation. Note, that for the generation of exact solutions $C_1=0$.\\

\subsection*{Improved AIM}

\par One unappealing feature of the recursion relations in equations (\ref{eqn:nthiter}) is that at each iteration one must take the derivative of the $s$ and $\lambda$ of the previous iteration. This can slow down the numerical implementation of the AIM and also lead to problems with numerical precision. To circumvent these issues we have developed an improved version of the AIM which bypasses the need to take derivatives at each step. This greatly improves both the accuracy and speed of the method. We expand the $\lambda_n$ and $s_n$ in a Taylor series around the point which the AIM is performed, $\xi$: 
\bea
\lambda_n(\xi)&=&\sum_{i=0}^{\infty}c_n^i(x-\xi)^i,\\
s_n(\xi)&=&\sum_{i=0}^{\infty}d_n^i(x-\xi)^i,
\eea
where the $c_n^i$ and $d_n^i$ are the $i$-th Taylor coefficient's of $\lambda_n(\xi)$ and $s_n(\xi)$ respectively. Substituting these expressions into equations \ref{eqn:nthiter} leads a set of recursion relations for the coefficients:
\bea\label{eqn:coeffiterlam}
c_n^i&=&(i+1)c_{n-1}^{i+1}+d_{n-1}^{i}+\sum_{k=0}^{i}c_0^kc_{n-1}^{i-k},\\
d_n^i&=&(i+1)d_{n-1}^{i+1}+\sum_{k=0}^id_0^kc_{n-1}^{i-k}. \label{eqn:coeffiters}
\eea
In terms of these coefficients the ``quantization condition'' (\ref{eqn:quantcond}) can be reexpressed:
\beq
d_n^{0} c_{n-1}^{0}-d_{n-1}^{0}c_n^0=0
\eeq
and thus we have reduced the AIM into a set of recursion relations which no longer requires the derivative operator. 

\par Observing that the right hand side of equations (\ref{eqn:coeffiterlam}) and (\ref{eqn:coeffiters}) involve terms of order at most $n-1$, one can recurse these equations until only $c_0^i$ and $d_0^i$ terms remain (that is the coefficients of $\lambda_0$ and $s_0$ only). However for large numbers of iterations, due to the large number of terms, such expressions become impractical to compute. We avert this combinatorial problem by beginning at the $n=0$ stage and calculating the $n+1$ coefficients sequentially until the desired number of recursions is reached. Since the quantisation condition only requires the $i=0$ term, at each iteration $n$ we only need to determine coefficients with $i<N-n$, where $N$ is the maximum number of iterations to be performed. 

\par The QNMs that we calculate in this paper will be determined using this improved asymptotic iteration method. Indeed in a previous work of ours for spheroidal harmonics \cite{Cho:2009wf}, we found that the AIM was somewhat slower than the continued fraction method. However, the {\it Mathematica} code implementing this new version of the AIM is found to be on a par with the continued fraction approach; these can be downloaded from \cite{homepage}.  

\begin{table}[t]
\caption{\small\sl Quasinormal modes to 4 decimal places for gravitational perturbations ($
\sigma = -3$) where the fifth column is taken from reference \cite{Iyer:1986nq}. Note that the imaginary part of the $n=0$, $\ell=2$ result in \cite{Iyer:1986nq} has been corrected to agree with reference \cite{Leaver:1985ax} (see $[*]$). Note, if the number of iteration in the AIM is increased to say about $50$ then we find agreement with reference \cite{Leaver:1985ax} accurate to $6$ significant figures. }
\label{tab:1}
\begin{ruledtabular}
\begin{tabular}{ccccc}
$\ell$ & $n$ & $\omega_{Leaver}$  & $\omega_{AIM} ~{\rm(after~15~iterations)} $ & $
\omega_{WKB}$ \\ \hline
2 & 0 & 0.3737 - 0.0896 i [*]  & 0.3737 - 0.0896 i & 0.3732 - 0.0892 i \\
&&&($<$0.01\%)($<$0.01\%)& (-0.13\%)(0.44\%) [*] \\
& 1 & 0.3467 - 0.2739 i   & 0.3467 - 0.2739 i & 0.3460 - 0.2749 i \\
&&&($<$0.01\%)($<$0.01\%)& (-0.20\%)(-0.36\%) \\
& 2 & 0.3011 - 0.4783 i  & 0.3012 - 0.4785 i & 0.3029 - 0.4711 i \\
&&&(0.03\%)(-0.04\%)& (0.60\%)(1.5\%) \\
& 3 &0.2515 - 0.7051 i & 0.2523 - 0.7023 i & 0.2475 - 0.6703 i \\ 
&&&(0.32\%)(0.40\%)& (-1.6\%)(4.6\%) \\
3 & 0 &0.5994 - 0.0927 i  & 0.5994 - 0.0927 i & 0.5993 - 0.0927 i \\
&&&($<$0.01\%)($<$0.01\%)& (-0.02\%)(0.0\%) \\
& 1 & 0.5826 - 0.2813 i   & 0.5826 - 0.2813 i & 0.5824 - 0.2814 i \\
&&&($<$0.01\%)($<$0.01\%)& (-0.03\%)(-0.04\%) \\
& 2 &0.5517 - 0.4791 i & 0.5517 - 0.4791 i & 0.5532 - 0.4767 i \\
&&&($<$0.01\%)($<$0.01\%)& (0.27\%)(0.50\%) \\
& 3 &0.5120 - 0.6903 i& 0.5120 - 0.6905 i & 0.5157 - 0.6774 i \\
&&&($<$0.01\%)(-0.03\%)& (0.72\%)(1.9\%) \\
& 4 &0.4702 - 0.9156 i & 0.4715 - 0.9156 i & 0.4711 - 0.8815 i \\
&&&(0.28\%)($<$0.01\%)& (0.19\%)(3.7\%) \\
& 5 &0.4314 - 1.152 i & 0.4360 - 1.147 i & 0.4189 - 1.088 i \\
&&&(1.07\%)(0.43\%)& (-2.9\%)(5.6\%) \\
4 & 0 & 0.8092 - 0.0942 i  & 0.8092 - 0.0942 i & 0.8091 - 0.0942 i \\
&&&($<$0.01\%)($<$0.01\%)& (-0.01\%)(0.0\%) \\
& 1 & 0.7966 - 0.2843 i & 0.7966 - 0.2843 i & 0.7965 - 0.2844 i \\
&&&($<$0.01\%)($<$0.01\%)& (-0.01\%)(-0.04\%) \\
& 2 & 0.7727 - 0.4799 i  & 0.7727 - 0.4799 i & 0.7736 - 0.4790 i \\
&&&($<$0.01\%)($<$0.01\%)& (0.12\%)(0.19\%) \\
& 3 &0.7398 - 0.6839 i & 0.7398 - 0.6839 i & 0.7433 - 0.6783 i \\
&&&($<$0.01\%)($<$0.01\%)& (0.47\%)(0.82\%) \\
& 4 &0.7015 - 0.8982 i& 0.7014 - 0.8985 i & 0.7072 - 0.8813 i \\
&&&(-0.01\%)(-0.03\%)& (0.81\%)(1.9\%) 
\end{tabular}
\end{ruledtabular}
\end{table}

\section{Schwarzschild $\bf{d}$S}
\label{extra}

\par \textcolor{black}{The QNMs for Schwarzchild gravitational perturbations are presented in Table \ref{tab:1}; however, to further} justify the use of this method, it is instructive to consider some more general cases. As such, we shall now consider the Schwarzschild de Sitter (SdS) case, where we have the same WKB-like wave equation and potential as in equation (\ref{radtort}), but now 
\beq
f(r) = 1 - {2M\over r} - \Lambda {r^2\over 3}~,
\eeq
where $\Lambda >0$ is the cosmological constant. Interestingly, the choice of coordinates we use here leads to a simpler AIM solution, because there is no Coulomb power law tail; however, in the limit $\Lambda=0$ we recover the Schwarzschild results. Note that although it is possible to find an expression for the maximum of the potential in equation (\ref{radtort}), for the SdS case, it is the solution of a cubic equation, which for brevity we refrain from presenting here. In our AIM code \cite{homepage} we use a numerical routine to find the root to make the code more general.

\par In the SdS case it is more convenient to change coordinates to $\xi = 1/r$ \cite{Moss:2001ga}, which leads to the following master equation (cf. equation (\ref{mastpot})):
\beq
\frac{d^2 \psi}{d\xi^2} + \frac{p'}{p} \frac{d\psi}
{d\xi} + \left[ \frac{\omega^2}{p^2} 
- { \ell (\ell + 1)+ (1 - s^2)\left(2 M \xi - (4-s^2){\Lambda\over 6\xi^2}\right)  \over p}\right]\psi = 0~,
\label{masterxi}
\eeq
where we have defined 
\beq
p= \xi^2 - 2M \xi^3 -\Lambda /3 \qquad\qquad \Rightarrow \qquad\qquad p' = 2\xi(1-3M\xi)~.
\eeq 
It may be worth mentioning that for SdS we can express \cite{Moss:2001ga}: 
\beq
e^{i\omega x} =  (\xi-\xi_1)^{{i\omega\over2 \kappa_1}} (\xi-\xi_2)^{{i\omega\over2 \kappa_2}} (\xi-\xi_3)^{{i\omega\over2 \kappa_3}}
\eeq
in terms of the roots of $f(r)$, where $\xi_1$ is the event horizon and $\xi_2$ is the cosmological horizon (and $\kappa_n$ is the surface gravity at each $\xi_n$). This is useful for choosing the appropriate scaling behaviour for QNM boundary conditions.

\par Based on the above equation an appropriate choice for QNMs is to scale out the divergent behaviour at the cosmological horizon:\footnote{Note that this is opposite to the case presented in reference \cite{Moss:2001ga}, where they define the QNMs as solutions with boundary conditions: $\psi(x)\propto e^{\mp i \omega x}$ as $x\to \pm \infty$, cf. equation (\ref{QNMdef}), for $e^{i \omega t}$ time dependence.}
\beq
\psi(\xi) = e^{i\omega x} u (\xi)~, 
\label{SdScale}
\eeq
which implies
\beq
p u'' + (p'- 2 i\omega)u' - \left[\ell(\ell+1)+ (1 - s^2)\left(2 M \xi - (4-s^2){\Lambda\over 6\xi^2} \right)\right]u=0 \,\,\, , 
\label{youeq}
\eeq
in terms of $\xi$. Furthermore, based on the scaling in equation (\ref{SdScale}), the correct  QNM condition at the horizon $\xi_1$ implies: 
\beq
u(x)=  (\xi-\xi_1)^{-{i\omega\over\kappa_1}}\chi(x)~,
\eeq
where $\kappa_1$ is the surface gravity at the event horizon $\xi_1$:
\beq
\kappa_1 = \left.\frac 1 2 {d f\over dr}\right|_{r\to r_1} 
=M \xi_1^2 - \frac 1 3 {\Lambda\over \xi_1} \,\,\, , 
\eeq
with
$\xi_1=1/r_1$, where $r_1$ is the smallest real solution of $f(r) =0$, implying $p=0$.
The differential equation then takes the standard AIM form:
\bea
\chi''& = & \lambda_0(\xi)  \chi' + s_0(\xi) \chi~ ,
\eea
where
\bea
\lambda_0(\xi) &=& -\frac 1 p \left[p'- {2i\omega \over \kappa_1(\xi-\xi_1)} - 2 i\omega\right]~ , \\
s_0 (\xi) &= & \frac 1 p \left[\ell(\ell+1)+ (1-s^2) \left(2 M \xi - (4-s^2){\Lambda\over 6\xi^2} \right)\ +{i\omega\over \kappa_1(\xi-\xi_1)^2}\Big({i\omega\over\kappa_1} +1\Big)
+(p'- 2 i\omega) {i\omega\over \kappa_1(\xi-\xi_1)}\right]~,
\eea
where in Table \ref{tab:2} results are presented for SdS with $s=2$. 

\par For completeness in the Appendix we derive a set of three-term recurrence relations for the continued fraction method, valid for electromagnetic and gravitational perturbations ($s=1,2$). It may be worth mentioning that via the AIM we can treat the $s=0,1,2$ perturbations on an equal footing (for the scalar case the continued fraction method reduces to a five-term recurrence relation, see the Appendix).


\begin{table}[h]
\caption{\small\sl Quasinormal modes to 6 significant figures for Schwarzschild de Sitter gravitational perturbations ($s=2$) for $\ell=2$ and $\ell=3$ modes. We only present results for the AIM method, because the results are identical to those of the continued fraction method after a given number of iterations (in this case $50$ iterations for both methods). The $n=1,2$ modes can be compared with the results in \cite{Zhidenko:2003wq} for $s=2$.}
\label{tab:2}
\begin{ruledtabular}
\begin{tabular}{|r|r|r|r|}
$\Lambda ~(\ell=2)$ & $n=1$ & $n=2$  & $n=3$ \\ 
\hline
0 		& 	0.373672 -  0.0889623 i	& 	0.346711 -  0.273915 i	&  	0.301050 -  0.478281 i	\\
0.02 	& 	0.338391 -  0.0817564 i	&   0.318759 -  0.249197 i	&  	0.282732 -  0.429484	 i\\
0.04 	& 	0.298895 -  0.0732967 i	&  	0.285841 -  0.221724 i	&   	0.259992 -  0.377092	 i\\
0.06		& 	0.253289 -  0.0630425 i	&	0.245742 -  0.189791 i	& 	0.230076 -  0.319157	 i\\ 
0.08  	& 	0.197482 -  0.0498773 i	&  	0.194115 -  0.149787 i	&   	0.187120 -  0.250257	 i\\
0.09 	& 	0.162610 -  0.0413665 i	&   0.160789 -  0.124152 i	&   0.157042 -  0.207117	 i\\
0.10 	& 	0.117916 -  0.0302105 i	& 	0.117243 -  0.0906409 i&   	0.115876 -  0.151102	 i\\
0.11 	& 	0.0372699 -  0.00961565 i	& 	0.0372493 -  0.0288470 i	&  	0.0372081 -  0.0480784 i	\\
\end{tabular}
\hspace{2.5cm}
\begin{tabular}{|r|r|r|r|}
$\Lambda ~(\ell=3)$ & $n=1$ & $n=2$  & $n=3$ \\ 
\hline
0 		& 	0.599443 -  0.0927030 i	& 	0.582644 -  0.281298	 i	&  	0.551685 -  0.479093	 i\\
0.02 	& 	0.543115 -  0.0844957 i	&   0.530744 -  0.255363 i 		&  	0.507015 -  0.432059 i	\\
0.04 	& 	0.480058 -  0.0751464 i	&  	0.471658 -  0.226395 i		&   	0.455011 -  0.380773	 i\\
0.06		& 	0.407175 -  0.0641396 i	&	0.402171 -  0.192807 i		& 	0.392053 -  0.322769	 i\\ 
0.08  	& 	0.317805 -  0.0503821 i	&  	0.315495 -  0.151249 i		&   	0.310803 -  0.252450	 i\\
0.09 	& 	0.261843 -  0.0416439 i	&   0.260572 -  0.124969 i 		&   	0.257998 -  0.208412	 i\\
0.10 	& 	0.189994 -  0.0303145 i	& 	0.189517 -  0.0909507  i		&   	0.188555 -  0.151609	 i\\
0.11 	& 	0.0600915 -  0.00961888 i &	0.0600766 -  0.0288567 i	&  	0.0600469 -  0.0480945  i\\
\end{tabular}
\end{ruledtabular}
\end{table}


\section{Schwarzschild A$\bf{d}$S}
\label{extraA}

\par There are various approaches to finding QNMs for the SAdS case (an eloquent discussion is given in the  appendix of reference \cite{Berti:2003ud}). One approach is that of Horowitz \& Hubeny \cite{Horowitz:1999jd}, which uses a series solution chosen to satisfy the SAdS QNM boundary conditions. This method can easily be applied to all perturbations ($s=0,1,2$). The other approach is to use the Frobenius method of Leaver \cite{Leaver:1985ax}, but instead of developing a continued fraction, the series must satisfy a boundary condition, such as Dirichlet, at infinity \cite{Moss:2001ga}.

\par The AIM does not seem easy to apply to metrics where there is an asymptotically anti-de Sitter background, because for general spin, $s$, the potential at infinity is a constant and hence would include a combination of ingoing and outgoing waves, leading to a sinusoidal dependence \cite{Cardoso:2001bb}. However, for the scalar spin zero ($s=0$) case, the potential actually blows up at infinity and is effectively a bound state problem. In this case the AIM can easily be applied as we show below.

\par Let us consider the scalar wave equation in SAdS spacetime, where $\Lambda= - 3/R^2$, and $R$ is the AdS radius. \textcolor{black}{The master equation takes the same form as for the graviational case \ref{radtort} except that the potential becomes:
\begin{equation}
V=\left(1-\frac{2}{r}+r^{2}\right)
\left(\frac{2}{r^{3}}+2\right)=\frac{2(r-1)(r^{2}+r+2)(r^{3}+1)}{r^{4}} \,\,\, . 
\end{equation}
}Here for simplicity we have taken the AdS radius $R=1$, the mass of the black hole $M=1$, and the angular momentum number $l=0$. Hence, the horizon radius $r_{+}=1$. Thus, with this choice we can compare with the data in Table 3.2 on page 37 of reference \cite{Cardoso:2003pj} (see Table \ref{tab:3} below).

\par To implement the AIM we first look at the asymptotic behavior of $\psi$. One could obtain this by using the WKB approximation:
\begin{equation}
\psi\sim\frac{1}{\left|\omega^{2}-V\right|^{1/4}}e^{\pm i\int\frac{\sqrt{\omega^{2}-V}}{1-\frac{2}{r}+r^{2}}dr} \,\,\, , 
\end{equation}
As $r\rightarrow r_{+}=1$, the potential $V$ goes to zero. In addition,
\begin{eqnarray}
  \textcolor{black}{x}&=&\int\frac{rdr}{(r-1)(r^{2}+r+2)}\sim \frac{1}{4}{\rm ln}(r-1)+\cdots\\
\left(\omega^{2}-V\right)^{-1/4}&\sim&\frac{1}{\sqrt{\omega}}+\frac{4}{\left(\sqrt{\omega}\right)^{5}}(r-1)+\cdots\\
\int dr\frac{\sqrt{\omega^{2}-V}}{1-\frac{2}{r}+r^{2}}&\sim&\int dr\left[\frac{\omega}{4(r-1)}+\cdots\right]\sim\frac{\omega}{4}{\rm ln}(r-1)+\cdots\\
\psi&\sim&e^{\pm i\left[\frac{\omega}{4}{\rm ln}(r-1)\right]}\sim (r-1)^{\pm i\omega/4}\sim \left(1-\frac{1}{r}\right)^{\pm i\omega/4} \,\,\, . 
\end{eqnarray}
For QNMs we choose the ``out-going" (into the black hole) boundary condition. That is,
\begin{equation}
  \psi\sim e^{-i\omega \textcolor{black}{x}}\sim \left(1-\frac{1}{r}\right)^{-i\omega/4} \,\,\, . 
\end{equation}

\par On the other extreme of our space, $r\rightarrow\infty$, the potential goes to infinity. This is a crucial difference from the case of gravitational perturbations. In that case, the potential goes to a constant. 
However, in the scalar case, as $r\rightarrow\infty$,
\begin{eqnarray}
  \textcolor{black}{x}&\sim&-\frac{1}{r}+\cdots\\
\left(\omega^{2}-V\right)^{-1/4}&\sim&\left(\frac{1}{r}\right)^{1/2}+\cdots\\
\int dr\frac{\sqrt{\omega^{2}-V}}{1-\frac{2}{r}+r^{2}}&\sim&-i\sqrt{2}\ {\rm ln}\left(\frac{1}{r}\right)+\cdots\\
\psi&\sim&\left(\frac{1}{r}\right)^{\frac{1}{2}\pm\sqrt{2}} 
\end{eqnarray}
and to implement the Dirichlet boundary condition, we take
\begin{equation}
\psi\sim\left(\frac{1}{r}\right)^{\frac{1}{2}+\sqrt{2}} \,\,\, .
\end{equation}

\par For the AIM one possible choice of variables is
\begin{equation}
\xi=1-\frac{1}{r}
\end{equation}
and we see that to accommodate the asymptotic behaviour of the wavefunction we should take
\begin{equation}
\psi=\xi^{-i\omega/4}(1-\xi)^{\sqrt{2}+\frac{1}{2}}\chi \,\,\, . 
\end{equation}
Finally, after some calculation we find the scalar perturbation equation to be:
\bea
\chi''& = & \lambda_0(\xi)  \chi' + s_0(\xi) \chi~ ,
\eea
where
\begin{eqnarray}
\lambda_{0}&=&-\frac{-i\omega q+2[-4+2(9+4\sqrt{2})\xi-(21+10\sqrt{2})\xi^{2}+4(2+\sqrt{2})\xi^{3}]}
{2\xi q} \,\,\, , \nonumber\\ \\
s_{0}&=&\frac{1}{16\xi q^{2}}\bigg(4i\omega[9+8\sqrt{2}-2(7+5\sqrt{2})\xi+(6+4\sqrt{2})\xi^{2}]q +\omega^{2}(-1+\xi)^{2}(-40+41\xi-20\xi^{2}+4\xi^{3})\nonumber\\
&&\ \ -4[4-5\xi+2\xi^{2}][-8(3+2\sqrt{2})+8(10+7\sqrt{2})\xi-(91+64\sqrt{2})\xi^{2}+(34+24\sqrt{2})\xi^{3}]\bigg) ~,\nonumber\\
\end{eqnarray}
and $q=(-4+9\xi-7\xi^{2}+2\xi^{3})$. Using the AIM we find the results presented in Table \ref{tab:3} below, which are discussed in the next section.

\begin{table}[h]
\caption{\small\sl Comparison of the first few quasinormal modes to 6 significant figures for Schwarzschild anti de Sitter scalar perturbations ($s=0$) for $\ell=0$ modes with $r_+=1$. The second column corresponds to data \cite{Cardoso:2003pj} using the Horowitz \& Hubeny method \cite{Horowitz:1999jd}, while the third column is for the AIM using 70 iterations. Note the mismatch for real part of the $n=3$ mode in \cite{Cardoso:2003pj},  see [*] below; we have confirmed this using the {\it Mathematica} notebook provided in \cite{Berti:2009kk} .}
\label{tab:3}
\begin{tabular}{|r|c|c|}
\hline
$n$ & HH method & AIM \\ 
\hline
0	& 	2.7982 - 2.6712 i	& 		2.79823 -2.67121 i  	\\
1	& 	4.75849 - 5.03757 i	& 		4.75850 -5.03757 i  	\\
2	& 	6.71927 - 7.39449 i	& 		6.71931 -7.39450  i 	\\
3	&   8.68223[*] - 9.74852 i	&   8.68233 -9.74854  i 	\\
4	& 	10.6467 - 12.1012 i	& 		10.6469 -12.1013 i  	\\
5	& 	12.6121 - 14.4533 i	& 		12.6125 -14.4533 i  	\\
6	& 	14.5782 - 16.8049 i	& 		14.5788 -16.8050 i  	\\
7	& 	16.5449 - 19.1562 i	& 		16.5457 -19.1563 i  	\\
\hline
\end{tabular}
\end{table}

\section{Results and Discussion}
\label{results}

\par We first applied the AIM to the Schwarzschild background defined in Section \ref{schwarz}, setting 
$M=1$ and choosing $\xi = \xi_0$ (the black hole maximum), which typically leads to the fastest 
converging results. In Table \ref{tab:1} the QNMs for graviton perturbations using the AIM are 
compared to some other methods.  These results show that after fifteen iterations the AIM is in 
good agreement with Leaver's method to four significant figures, with the disagreement 
becoming most pronounced for the lowest $\ell$ modes with larger values of $n$. We also 
found that by increasing the number of iterations the AIM can be made to agree exactly with 
the results of Leaver's continued fraction method \cite{Leaver:1985ax} to any number of decimal places. 

\par For completeness we have also included results from an approximate semi-analytic 3rd order WKB 
method \cite{Iyer:1986nq} (more accurate semi-analytic results with better agreement to 
Leaver's method can be obtained by extending the WKB method to 6th order \cite{Konoplya:2003ii}). 
In this regard, it may be worth mentioning that a different semi-analytic approach has recently been discussed by Dolan and Ottewill \cite{Dolan:2009nk}, which has the added 
benefit of easily being extended to any order in their perturbative scheme.

\par We also discussed spin $s=0,1,2$ in SdS (see Table \ref{tab:2}) where results to six significant figures are presented for the graviton ($s=2$) case. Identical results were generated by the AIM and  continued fraction method, both after 50 iterations. Though results are presented for $n = 1, 2$ and $3$; $\ell = 2$ and $3$ modes only, the AIM is robust enough to be applied to any other case; where like for the $\Lambda = 0$ case, agreement with other methods in more extreme parameter choices would only require further iterations. For comparison with the AIM we have also used the continued fraction method, and for completeness we presented a derivation in the appendix (for spin $s=1,2$). 

\par As far as we are aware, only reference \cite{Zhidenko:2003wq} (who used a semi-analytic WKB approach) has presented tables for  general spin fields for the SdS case. We have also compared our results to those in \cite{Zhidenko:2003wq} for the $s=0,1$ cases and find identical results (to a given accuracy in the WKB method). As briefly discussed in the appendix another benefit of the AIM is that no {\it Gaussian Elimination} procedure is required; unlike for continued fractions on Reissner-N\"ordtrom \cite{Leaver:1990zz} or higher-dimensional Schwarzschild backgrounds \cite{Zhidenko:2006rs}, for example. All that is necessary is to factor out the correct asymptotic behaviour at the horizon(s) and infinity (we showed this for higher-dimensional angular spheroids in \cite{Cho:2009wf}). Of course, there is no need to find the maximum of the black hole potential for continued fractions, unlike with the AIM.

\par The reader might wonder about approximate results for cases where Poschl-Teller approximations can be used, such as for Schwarzschild and SdS backgrounds, e.g., see  \cite{Ferrari:1984zz, Moss:2001ga}. In fact when the black hole potential can be modeled by a Scarf like potential the AIM can be used to find the eigenvalues exactly \cite{OzerRoy} and hence the approximate QNMs. Indeed the authors of reference \cite{OzerRoy} also independently suggested the idea of using the AIM to find black hole QNMs from exact potentials numerically (based on previous comments by one of the present authors \cite{ChoLin}). On a related note the large $\ell$ eikonial limit (e.g., for $\Lambda=0$ see \cite{Iyer:1986nq}) is also easily confirmed numerically as we have verified using the continued fraction method \cite{homepage}.

\par The SAdS case is a little more subtle, and we briefly discussed some preliminary results for a spin zero field (the $l=0$, $r_+=1$ case) in Table \ref{tab:3}. In the scalar case the WKB form of the potential is essentially a bound state potential (it blows up at infinity) and allowed us to apply the AIM numerically. For other spins the potential goes to a constant at infinity and thus would be a combination of ingoing and outgoing waves. This implies a sinusoidal boundary condition at infinity (e.g., see \cite{Cardoso:2001bb}) which is a little awkward to use with the AIM and somewhat out of the scope of this article. 

\par In summary, we have demonstrated how the AIM can also be applied to radial 
QNMs and not just to spheroidal eigenvalue problems \cite{Barakat:2006,Cho:2009wf}. Given 
the fact that the AIM can be used in both the radial and angular wave equations \cite{Cho:2009wf} we expect no 
problems in obtaining QNMs for Kerr or Kerr-dS black holes for example.  Furthermore, a perusal of the work on the AIM (e.g., see references \cite{Ciftci:2003,OzerRoy,Barakat:2006} and references therein) indicates that there are still some mathematical problems/questions of numerical refinement (such as the improved AIM we discussed), which was another reason for bringing this method to light.

\par As for future work, it remains to be seen if the AIM can be tailored to handle asymptotic QNMs (see reference \cite {Nollert:1993zz} for an adapted version of the continued fraction method). However, given the close relationship between the AIM and the exact WKB approach \cite{Matamala:2007} it might be possible to adapt the AIM to find asymptotic QNMs \cite{Motl,Andersson} numerically or even semi-analytically. Also, there are still issues for the SAdS case, but we hope to report on more general results for this in the near future.

%

\acknowledgments

\par HTC was supported in part by the National Science Council of the Republic of China 
under the Grant NSC 96-2112-M-032-006-MY3, and the National Science Centre for 
Theoretical Sciences. ASC would like to thank the Yukawa Institute for Theoretical Physics for 
their hospitality (under their GCOE visitor program). The work of JD was supported by the 
Japan Society for the Promotion of Science (JSPS), under fellowship no. P09749.

\appendix*
\section{Continued fraction method for S$\bf{d}$S with $s=1,2$}
\label{CFM}

\par For completeness and for the purpose of making an accurate comparison with the AIM method, for the SdS case we present some the steps that lead to the continued fraction method. As far as we are aware the actual recursion coefficients themselves do not seem to appear in the literature (e.g., see \cite{Moss:2001ga}). 

The correct scaling for QNM boundary conditions in SdS implies
\beq
\psi(x)=  e^{i\omega x} (\xi-\xi_1)^{-{i\omega\over\kappa_1}}\chi(x)~,
\eeq
where we can expand $\chi$ as a power series:
\beq
\chi = \sum_{n=1}^\infty a_n y^n~, \qquad y= {\xi -\xi_1 \over \xi_2 -\xi_1 } \qquad \Rightarrow \qquad \xi = (\xi_2-\xi_1) y + \xi_1 ~,
\eeq
where $a_0=1$. Note that the form of $y$, above, depends on $(\xi_2 -\xi_1)$ in order to give a converging solution in the Frobenius solution. It then follows that
if we define
$
\varphi = \sum_{n=1}^\infty a_n y^{n-{i\omega\over\kappa_1}}
$
then
\beq
\psi(x)=  e^{i\omega x} (\xi_2-\xi_1)^{-{i\omega\over\kappa_1}}\varphi(x)~,
\eeq
and hence the $(\xi_2 -\xi_1)$ term factors out of the master equation (\ref{masterxi}). The rescaled  equation for $\varphi$ then takes an identical form to equation (\ref{youeq}):
\beq
\xi^2 p \varphi'' + \xi^2(p'- 2 i\omega)\varphi' - \left[\ell(\ell+1)\xi^2+ (1 - s^2)\left(2 M \xi^3 - (4-s^2){\Lambda\over 6} \right)\right]\varphi=0 \,\,\, , 
\label{5term}
\eeq
where for general integer spin we multiplied both sides of the equation by $\xi^2$. We could then expand $\varphi$ as a power series where
\beq
 \varphi' = \sum_{n=1}^\infty a_{n}\bigg(n- {i\omega \over \kappa}\bigg)y^{n-{i\omega\over\kappa_1}-1}(\xi_2-\xi_1)^{-1}~,
 \qquad
 \varphi' = \sum_{n=1}^\infty a_{n}\bigg(n- {i\omega \over \kappa}\bigg)\bigg(n- {i\omega \over \kappa}-1\bigg)y^{n-{i\omega\over\kappa_1}-2}(\xi_2-\xi_1)^{-2}~,
\eeq
where we should also power expand $\xi^2 p$ and $\xi^2 p'$ in a power series about $\xi_1$ (for a detailed discussion of this point see \cite{Giammatteo:2005vu}). 

\par It is straightforward to see that this will lead to a five-term recurrence relation, because $\xi^2 p \sim {\cal O}(y^5)$, with $p(\xi_1)=0$, for the $u''$ term. This would then require the use of two {\it Gaussian Eliminations} \cite{Leaver:1990zz} to reduce it to a three-term recurrence relation. Note, in the Reissner-N\"ordtrom-(A)dS case, even the axial graviton perturbations reduce to a four-term recurrence relation \cite{Berti:2003ud}.

\par Thus, in order to obtain a set of three-term recurrence relations (which are simpler to deal with as no {\it Gaussian Eliminations} are required) we consider only spin $s=1,2$. This drops the last constant term in equation (\ref{5term}) and allows us to divide out by a factor of $\xi^2$. Then, because $p \sim {\cal O}(y^3)$ for the $u''$ term, this leads to a three-term recurrence relation. After substituting these expansions into equation (\ref{5term}) and equating coefficients we obtain a set of recurrence coefficients:
\bea
\alpha_n &=& 2(\xi_2-\xi_1)^{-1} (n- {i\omega \over \kappa}+1) \Big[ (n- {i\omega \over \kappa}+1)\xi_1 (1-3M\xi_1) - i\omega \Big] \nn
\beta_n &=&-\ell (\ell+1) - 2M(1-s^2)\xi_1 + (1-6M\xi_1) (n- {i\omega \over \kappa})(n- {i\omega \over \kappa}+1) \\
\gamma_n &=& -2M(\xi_2-\xi_1)(n- {i\omega \over \kappa}+s)(n- {i\omega \over \kappa}-s) \nonumber~,
\eea
which satisfy the following recurrence relations:
\bea 
\alpha_0 a_1 + \beta_0 a_0 &=& 0 \\
\alpha_n a_{n+1} + \beta_n a_n +\gamma_n a_{n-1} &=&0~.
\eea
The exterior eigenvalue problem (QNMs) then leads to a continued fraction \cite{Leaver:1985ax}:
\beq
0= \beta_0 - {\alpha_0\gamma_1\over \beta_1 -}{\alpha_1\gamma_2\over \beta_2 -}\dots~,
\eeq
where this equation can be solved for numerically and allows us to check the results with the AIM, see \cite{homepage}. In all cases we find perfect agreement between the two results (to a given precision) for roughly the same number of iterations in both the AIM and continued fraction. Note that in the recurrence coefficients  above, the overall factors of $(\xi_1-\xi_2)$ cancel out in the continued fraction.

%
%

\end{document}